\documentclass[11pt,a4paper]{article}
\usepackage[hyperref]{emnlp2020}
\usepackage{times}
\usepackage{latexsym}
\usepackage{graphicx}

\usepackage{booktabs}
\usepackage{multirow}
\usepackage{siunitx}
\usepackage[normalem]{ulem}
\usepackage{subfig}
\usepackage{caption}

\newcommand{\bertsliceaware}{BERT-SA}
\newcommand{\bertsliceawarerandom}{BERT-SA-R}

\usepackage{microtype}

\aclfinalcopy % Uncomment this line for the final submission
\title{Slice-Aware Neural Ranking}

\author{Gustavo Penha \\
  TU Delft \\
  \texttt{g.penha-1@tudelft.nl} \\\And
  Claudia Hauff \\
  TU Delft \\
  \texttt{c.hauff@tudelft.nl} \\}

\begin{document}
\maketitle
\begin{abstract}
Understanding when and why neural ranking models fail for an IR task via error analysis is an important part of the research cycle. Here we focus on the challenges of (i) identifying categories of \emph{difficult} instances (a pair of question and response candidates) for which a neural ranker is ineffective and (ii) improving neural ranking for such instances. To address both challenges we resort to \emph{slice-based learning}~\cite{chen2019slice} for which the goal is to improve effectiveness of neural models for slices (subsets) of data. We address challenge (i) by proposing different \emph{slicing functions} (SFs) that select slices of the dataset---based on prior work we heuristically capture different failures of neural rankers. Then, for challenge (ii) we adapt a neural ranking model to learn slice-aware representations, i.e. the adapted model learns to represent the question and responses differently based on the model's prediction of which slices they belong to. Our experimental results\footnote{The source code and data are available at \url{https://github.com/Guzpenha/slice_based_learning}.} across three different ranking tasks and four corpora show that slice-based learning improves the effectiveness by an average of 2\% over a neural ranker that is not slice-aware.
\end{abstract}

\section{Introduction}
\begin{figure}[]
    \centering
    \includegraphics[width=.45\textwidth]{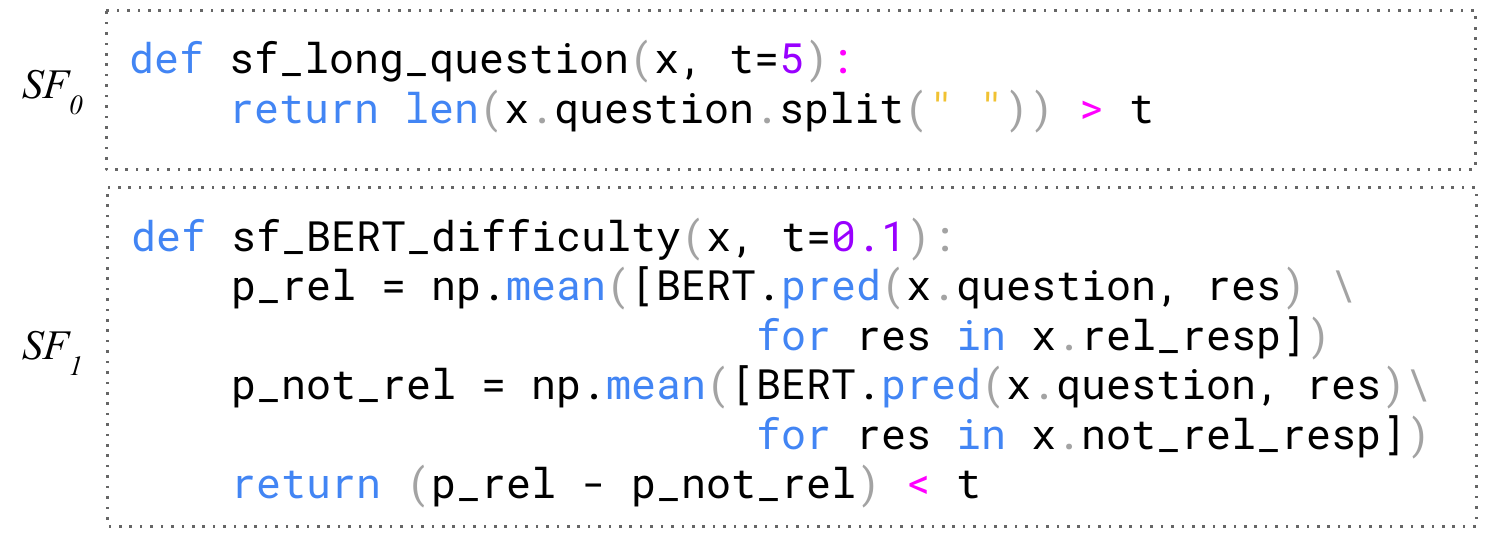}
    \setlength{\abovecaptionskip}{-1pt}
    \setlength{\belowcaptionskip}{-15pt}
    \caption{Examples of slicing functions (SFs) to capture subsets of difficult tuples of question and response list. The SFs also have access to relevance labels for the training set, as they are not required at test time by the slice-aware neural ranker. $SF_0$ uses the question length as a proxy for question complexity, and $SF_1$ calculates how distinguishable relevant and non-relevant responses are based on BERT predictions.
    }
    \label{fig:slicing_functions}
    % \vspace{-0.8cm}
\end{figure}

% quick introduction to IR
Retrieving text for a given information need is a fundamental task in Information Retrieval (IR). For a long time neural networks failed to convincingly outperform traditional term matching approaches with pseudo-relevance feedback, e.g. RM3~\cite{abdul2004umass}, for text retrieval tasks including the classic adhoc retrieval task~\cite{Yang:2019:CEH:3331184.3331340}. However, with recent breakthroughs in natural language processing (NLP), neural approaches---prominently BERT~\cite{devlin2019bert}---are achieving state-of-the-art effectiveness across a range of text retrieval tasks~\cite{yang2019simple,nogueira2019passage}.

% challenge and Slice-based-Learning
Understanding when and why retrieval models fail is an important part of the research cycle. Even tough we have clues about the failures of neural rankers---obtained for instance by the study of question performance prediction~\cite{he2006query}, diagnostic datasets~\cite{camara2020diagnosing} and error analysis~\cite{wu2019errudite}---automatically identifying difficult instances (tuples of question and response list) and improving the effectiveness of models for such difficult instances are still open challenges. We consider here difficult instances to be question and responses for which a given neural ranker retrieval effectiveness is below the average. A recent approach, referred to as \emph{slice-based learning}~\cite{chen2019slice}, has been proposed to identify and improve the effectiveness of subsets of data (so-called \emph{slices}), as opposed to focusing on all data equally. The core idea is that a slice-aware neural model will represent instances differently depending on the slices of data they come from. Slice-based learning has been applied to computer vision and NLP tasks, with overall effectiveness improvements up to 3.5\%~\cite{chen2019slice} over a model that is not slice-aware.

% our contributions
In this paper we focus on the challenges of (i) detecting difficult instances for neural rankers and (ii) improving the retrieval effectiveness for such instances. We address the challenges by (i) creating slicing functions (SFs), i.e., functions that define whether an instance belongs to a slice which heuristically capture different errors of rankers (cf. Figure~\ref{fig:slice_aware_neural_ranker} for examples of SFs); and (ii) employing a slice-aware neural ranker, i.e., a neural ranker that learns to represent each instance differently based on its prediction of which slice the input belongs to (cf. Figure~\ref{fig:slice_aware_neural_ranker} for a diagram of the slice-aware neural ranker). Our main research questions are the following two. \textbf{RQ1:} To what extent can slice-based learning improve neural ranking models? \textbf{RQ2}: What are the underlying reasons for the effectiveness of slice-based learning?

%  our results.
Our experimental results on three different conversational tasks show that slice-based learning is beneficial to IR, showing positive evidence for RQ1. The gains are observed for both overall effectiveness and the effectiveness for slices of the data. Concerning RQ2, we evaluate to which extent the effectiveness gains observed for the slice-aware model come from the effect of ensemble learning~\cite{dietterich2002ensemble}, a direction not explored empirically by previous work~\cite{chen2019slice}. We find that, when using \emph{random} SFs we can also significantly improve upon a non slice-aware neural ranker. We note though that not all improvements of slice-based learning can be attributed to the effect of ensemble learning, and carefully implementing SFs is indeed advantageous.
\section{Slice-based Learning}
% \subsubsection*{Slice-based Learning}
Slice-based learning~\cite{chen2019slice} is an approach based on the engineering of SFs that capture slices of data. The SFs all follow the same format: they receive the instance as input (in our case a question and a list of candidate responses) and return a boolean variable indicating whether the instance belongs to the slice. Based on the SFs a neural model is adapted to improve the effectiveness of such slices of data, for example, by having a different set of weights for each slice. Training a different model for each slice, and combining their predictions is inefficient: training and maintaining a different neural ranking model for each slicing function amounts to a large number of parameters and an increased prediction time. As an efficient solution, \citet{chen2019slice} proposed Slice-Residual-Attention Modules (SRAMs), which is a slice-aware approach for neural models that shares parameters in a similar manner to multi-task learning~\cite{caruana1997multitask}.

\section{Slice-based learning for IR}

\begin{figure}[]
    \centering
    \includegraphics[width=.49\textwidth]{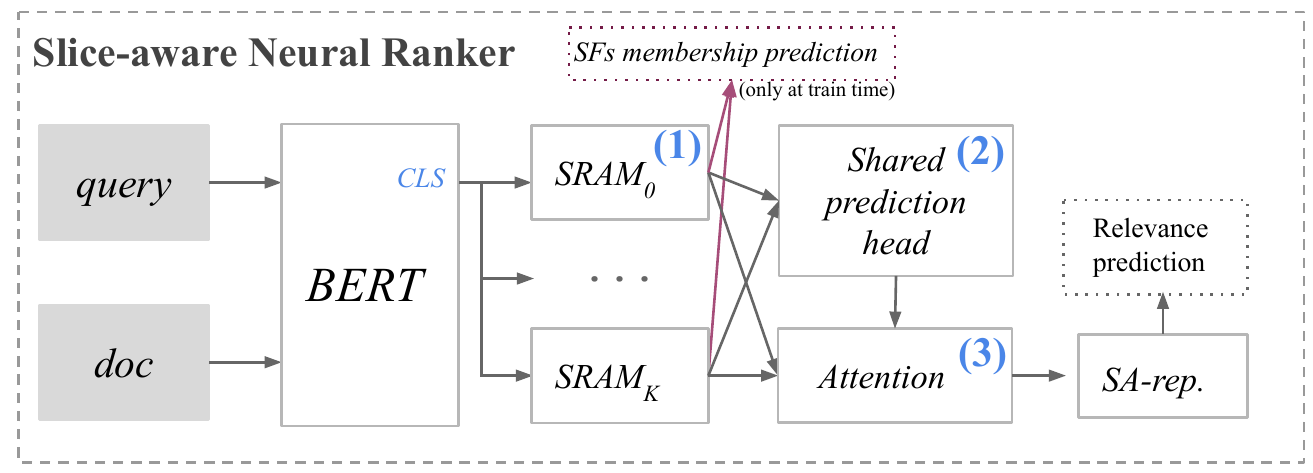}
    \setlength{\abovecaptionskip}{-10pt}
    \setlength{\belowcaptionskip}{-10pt}
    \caption{
    Overview of the slice-aware neural ranker. For each SF we define we have a SRAM module to learn slice-expert representations, that are then combined with an attention mechanism into a slice-aware representation.}
    \label{fig:slice_aware_neural_ranker}
    \vspace{-0.1cm}
\end{figure}

% \begin{figure*}[]
%     \centering
%     \includegraphics[width=1\textwidth]{img/slice_aware_neural_ranker_with_functions.pdf}
%     % \setlength{\abovecaptionskip}{-10pt}
%     % \setlength{\belowcaptionskip}{-15pt}
%     \caption{\textbf{Left:} examples of slicing functions (SFs) to capture subsets of difficult tuples of question and response list. $SF_0$ uses the question length as a proxy for question complexity, and $SF_1$ calculates how distinguishable relevant and non-relevant responses are based on BERT predictions. \textbf{Right:} overview of the slice-aware neural ranker. For each SF we define we have a SRAM module to learn slice-expert representations, that are then combined with an attention mechanism into a slice-aware representation.}
%     \label{fig:slice_aware_neural_ranker}
%     % \vspace{-0.1cm}
% \end{figure*}

We first introduce the SFs we defined to heuristically capture subsets of data containing different categories of errors, for which the effectiveness is lower than average, based on intuitions drawn from prior work (RQ1). We then introduce the random SFs we deploy to study the effect of ensemble learning in slice-based learning (RQ2). Finally we describe the slice-aware neural ranker.

\subsection{Slicing Functions}\label{section:slicing_functions}
We divide our SFs into two categories: those based only on the question text (question based) and those that uses both the question and the list of candidate responses (question-responses based). The relevance labels for the training instances are also inputs to the SFs, which are not required at inference time as the slice-aware neural ranker learns to predict slice-membership.

\subsubsection{Question-based SFs}\label{section:question_based}

\uline{Question Length (QL)}: the number of question terms is higher than the threshold $T_{QL}$. QL was shown to correlate negatively with the effectiveness of retrieval methods in adhoc retrieval~\cite{bendersky2009analysis}. Long questions (questions with high QL) provide a way of expressing complex information needs as opposed to short questions~\cite{phan2007understanding}. \uline{Context Length (CL)}~\footnote{This SF is only suited for QA tasks with multiple turns.}: the number of turns in the dialogue context is higher than the threshold $T_{CL}$. CL was shown to correlate negatively with model's effectiveness for the conversation response ranking task when using different neural rankers~\cite{tao2019multi}. \uline{Question Category (QC)}: question is about a certain semantic category, e.g. $QC=travel$ selects questions about travel. Knowing which topic a question belongs to can lead to retrieval effectiveness improvements, for instance by using federated search~\cite{INR-010}, intent-aware ranking~\cite{glater2017intent} or multi-task learning~\cite{liu2015representation}. Instances from different categories could display different effectiveness values, e.g. questions about \textit{physics} could be a potential difficult category. \uline{Question type (5W1H)}: a categorization into types of question (who, what, where, when, why, how), e.g. $5W1H=what$ selects \textit{what} questions. 5W1H has been used to inform dialogue management modules~\cite{han2013counseling}. The type of question can yield different models' effectiveness~\cite{kim2019probing}.

\vspace{-0.2cm}
\subsubsection{Question-Responses based SFs}\label{section:question_doc_based}

\uline{Question Response Term Match (QDTM)}: The number of words that appear in both the question and a relevant response is smaller than the threshold $T_{QDTM}$. The difference in vocabulary, i.e. lexical gap, between queries and documents has shown to be a problem in IR~\cite{lee2008bridging} and has to lead to remedies such as query expansion~\cite{voorhees1994query} and the use of neural ranking models for semantic matching~\cite{guo2019deep}. \uline{Responses Lexical Similarity (DLS)}: average TF-IDF similarity between the top-k most similar responses in the candidate list to the relevant response is higher than the threshold $T_{DLS}$. The amount of internal coherence, i.e. similarity between responses, has been used to predict query difficulty~\cite{he2008using}. The SFs can be easily extended for multiple relevant responses, e.g. by using the average or considering one representative relevant response.

\subsubsection{Random SFs}\label{section:random_based}
The random SF randomly samples $X\%$ of the training data, where $X$ is a hyperparameter.

\subsection{Slice-Aware Neural Ranker}

Figure~\ref{fig:slice_aware_neural_ranker} displays a diagram of the slice-aware neural ranker. Based on a backbone (BERT) that learns a representation of the question and response concatenation, the slice-aware neural ranker learns to \textcolor{blue}{(1)} predict how much each instance belongs to each of the $k$ slices or not (supervision is based on the boolean output of the $k$ SFs)\footnote{The model has an extra SF that all instances belong to, so every instance will always belong to at least to this slice.}; has $k$ slice expert representations with its own set of weights trained using a shared prediction head \textcolor{blue}{(2)} which predicts relevance for the question and response combination using only instances of the slice $k$; and \textcolor{blue}{(3)} combines all representations from the SRAMs using attention into a single slice-aware representation that is used to make the final relevance prediction. The SFs are only used during training and thus are not needed at inference time. This is an adaptation of SRAMs~\cite{chen2019slice}, and the backbone could be replaced by any other neural ranker.

\begin{table*}[ht!]
% \small
% \setlength{\abovecaptionskip}{-1pt}
\centering
\caption{Average of 5 runs for slice-based learning. Superscript $^{\dagger}$ denote statistically significant improvements over the baseline (BERT) where no slice-based learning is applied at 95\% confidence interval using Student's t-tests. Bold indicates the highest MAP for each dataset.}
\begin{tabular}{@{}llllll@{}}
\toprule
 & & \multicolumn{1}{c}{Dev} &  \multicolumn{3}{c}{Test}   \\
 \cmidrule(lr{0em}){3-3} \cmidrule(lr{1em}){4-6}
 & & \multicolumn{1}{l}{\multirow{2}{*}{MAP (std)}} & \multicolumn{1}{l}{\multirow{2}{*}{MAP (std)}} & \multicolumn{2}{l}{slice $\Delta$ MAP} \\ \cmidrule(lr{1em}){5-6} 
Dataset & Model & \multicolumn{1}{l}{} & & Avg. & Max. \\ \midrule
\multirow{3}{*}{\texttt{ANTIQUE}} & BERT & 0.853 (.026) & \multicolumn{1}{l|}{0.850 (.015)}  & \multicolumn{1}{c}{-} & \multicolumn{1}{c}{-} \\
 & \bertsliceawarerandom & 0.874 (.025)$^{\dagger}$ & \multicolumn{1}{l|}{0.877 (.005)$^{\dagger}$}  & 0.028 & 0.063 \\
 & \bertsliceaware &\textbf{0.878 (.024)$^{\dagger}$} & \multicolumn{1}{l|}{\textbf{0.883 (.005)$^{\dagger}$}}  & 0.035 & 0.112 \\ \midrule
\multirow{3}{*}{\texttt{MANtIS\_50}} & BERT & 0.655 (.006) & \multicolumn{1}{l|}{0.684 (.006)}  & \multicolumn{1}{c}{-} & \multicolumn{1}{c}{-} \\
 & \bertsliceawarerandom & 0.671 (.006)$^{\dagger}$ & \multicolumn{1}{l|}{\textbf{0.690 (.014)$^{\dagger}$}}  & 0.025 & 0.035 \\
 & \bertsliceaware &\textbf{0.702 (.006)$^{\dagger}$} & \multicolumn{1}{l|}{0.689 (.022)$^{\dagger}$}  & 0.025 & 0.034 \\ \midrule
\multirow{3}{*}{\texttt{MSDialog}} & BERT & 0.754 (.010) & \multicolumn{1}{l|}{0.830 (.002)}  & \multicolumn{1}{c}{-} & \multicolumn{1}{c}{-} \\
 & \bertsliceawarerandom & \textbf{0.815 (.009)$^{\dagger}$} & \multicolumn{1}{l|}{\textbf{0.840 (.011)$^{\dagger}$}}  & 0.028 & 0.084 \\
 & \bertsliceaware & 0.810 (.009)$^{\dagger}$ & \multicolumn{1}{l|}{0.818 (.010)}  & -0.004 & 0.067 \\ \midrule
\multirow{3}{*}{\texttt{Quora}} & BERT & 0.799 (.037) & \multicolumn{1}{l|}{0.819 (.008)}  & \multicolumn{1}{c}{-} & \multicolumn{1}{c}{-} \\
 & \bertsliceawarerandom & 0.819 (.035)$^{\dagger}$ & \multicolumn{1}{l|}{0.837 (.004)}  & 0.011 & 0.038 \\
 & \bertsliceaware & \textbf{0.834 (.034)$^{\dagger}$ }& \multicolumn{1}{l|}{\textbf{0.840 (.007)$^{\dagger}$}}  & 0.019 & 0.065 \\
 \bottomrule
\end{tabular}
\label{table:results}
% \vspace{-0.6cm}
\end{table*}

\section{Experimental Setup}
We employ four datasets and three retrieval tasks: \texttt{MSDialog}~\cite{qu2018analyzing} and \texttt{MANtIS}~\cite{penha2019introducing} for conversation response ranking, \texttt{Quora}~\cite{iyer2017first} for similar question retrieval and \texttt{ANTIQUE}~\cite{hashemi2019antique} for non-factoid question answering. We use the official train, validation and test sets provided by the datasets' creators. As a strong neural ranking baseline model we fine-tune BERT \footnote{\textit{bert-base-uncased} with default hyperparameters~\cite{Wolf2019HuggingFacesTS}.} for sentence classification, using the CLS token to predict whether the concatenation of a question and response is relevant or not, following recent research in IR~\cite{nogueira2019passage,yang2019simple}. Using 512 input tokens (larger inputs are truncated) and a batch size of 8 we train each model for 5 epochs. 

% We optimize BERT for point-wise learning, this way a instance is composed of a question and a single response here.
%  (if the model takes more than 32 hours for running 5 epochs of a dataset we stop at this time threshold)

When employing SRAMs~\cite{chen2019slice} with a BERT backbone for neural ranking using both the question-based and question-responses based SFs we refer to the model as \bertsliceaware. When using random SFs we refer to the model as \bertsliceawarerandom. For the SFs that have a threshold value (e.g., QL), we choose thresholds that select less than 50\% of the data to avoid selecting the majority of the training instances in each slice. For SFs that include a categorical value, e.g., question category (QC) \textit{physics}, we add one slice per category in the dataset. For the random SFs we create 10 different slices\footnote{Initial experiments varying the number of SFs showed a validation plateau around 10.} for which 50\% of randomly chosen instances from the training data belong to\footnote{Initial experiments varying revealed that only small percentages, less than 20\%, degraded the effectiveness.}. We train each model 5 times with different random seeds and report the test set effectiveness using Mean Average Precision (MAP). $\Delta$MAP indicates the difference between \bertsliceaware(\texttt{-R}) and BERT for the slices defined by the SFs.

%  (we study the correlation between slice size and effectiveness gains in \S\ref{section:results})

\section{Results}\label{section:results}
Let us first consider RQ1. We observe in Table~\ref{table:results} that with the exception of \texttt{MSDialog},~\bertsliceaware{} significantly improves over the baseline (BERT) for both the overall (column MAP) and per slice performance (column slice $\Delta$MAP). \textbf{This demonstrates that slice-based learning is useful for neural ranking, with gains up to 3.8\% overall and up to 13\% per slice in terms of MAP.}

% The results are shown in Table~\ref{table:slices_correlations}.
To better understand which features of a slice correlate the most with the observed gains from~\bertsliceaware{}, we study how three properties of the slices correlate with the slice $\Delta$MAP (i.e., the improvement over BERT): we consider (1) the size of the slice, (2) the classification accuracy of the slice-aware model to predict slice membership, and, (3) the BERT model effectiveness for each slice. The only property that has a statistically significant Pearson correlation (0.504 average for the different datasets) with MAP gains is the BERT baseline performance , suggesting that focusing on failures of neural ranking models (slices for which BERT has low effectiveness) when implementing SFs is effective.

To provide insights into the underlying reasons of the effectiveness of slice-based learning (RQ2), we replace the SFs that capture error categories with random SFs, i.e. \bertsliceawarerandom{}. We find that this model also has a significantly better effectiveness than the BERT baseline, with the exception of \texttt{Quora}. \textbf{This indicates that part of the gains provided by slice-based learning could be attributed to the effect of ensemble learning}, since each slice-aware representation is trained on random parts of the data and are then combined\footnote{Another potential reason for the success of slice-based learning could be the capacity obtained by the additional number of weights compared to the baseline (e.g. from 110M to 116M for \texttt{MANtIS}).}. We note however that the slice gains of \bertsliceaware{} are higher than \bertsliceawarerandom{} for \texttt{ANTIQUE} and \texttt{Quora} with statistical significance. This indicates that not all improvements of slice-based learning can be attributed to the effect of ensemble learning and carefully implementing SFs is advantageous.

\section{Conclusion}
In this paper we demonstrated that a slice-aware neural ranker is an effective approach to IR, increasing the effectiveness of rankers by margins up to 3.8\% overall and up to 13\% per slice in terms of MAP. As future work we plan to study slice-aware neural rankers that do listwise optimization---such a ranker could learn better representations particularly for SFs that uses several responses as input.

\section*{Acknowledgements}
This research has been supported by NWO projects SearchX (639.022.722) and NWO Aspasia (015.013.027).

\bibliographystyle{acl_natbib}
\bibliography{references}

\end{document}